# Nonlinearity-mediated soliton ejection from trapping potentials in nonlocal media

Fangwei Ye,[1] Yaroslav V. Kartashov,[2] Bambi Hu,[1,3] and Lluis Torner[2]

[1]*Department of Physics, Centre for Nonlinear Studies, and The Beijing-Hong Kong Singapore Joint Centre for Nonlinear and Complex Systems, Hong Kong Baptist University, Kowloon Tong, China*

[2]*ICFO-Institut de Ciencies Fotoniques, and Universitat Politecnica de Catalunya, Mediterranean Technology Park, 08860 Castelldefels (Barcelona), Spain*

[3]*Department of Physics, University of Houston, Houston, Texas 77204-5005, USA*

We address the properties of optical solitons in thermal nonlinear media with a local refractive index defect that is capable to trap solitons launched even close to the sample boundary despite the boundary-mediated forces that tend to deflect all beams toward the center of the sample. We show that while such forces become more pronounced with increasing the input beam power the defect can trap only light below a critical power above which solitons are ejected. The dynamics of soliton ejection and the subsequent propagation may be controlled.

In nonlocal nonlinear optical materials the nonlinear contribution to the refractive index created by an intense laser beam at each spatial point is affected by the intensity distribution in the entire transverse plane [1,2]. Such situation takes place when the nonlinearity mechanism in-volves diffusion processes or long-range interactions, and it has been encountered in many materials, including liquid crystals with reorientational nonlinearity [3,4], vapors where the atomic diffusion causes the transport of excitations away from the region of the light-matter interaction [5,6], and in materials exhibiting strong thermo-optic coefficients where heat redistribution causes notable refractive index modifications [7,8], among others. Note that besides optics, nonlocal nonlinearities arise in the description of a large variety of systems, vegetation patterns [9], population dynamics [10] and reaction-diffusion waves [11].

In optics nonlocality can have a profound effect on the properties of nonlinear excitations. For example, it may suppress beam collapse and transverse instabilities [12-14], or it may change the character of soliton interactions [15]. In materials where diffusion processes are involved, such as thermal media, boundary conditions may strongly impact the dynamics of soliton formation. Boundary effects in thermal media have been exploited for beam steering [16-19], generation of surface waves [20-22], and for the formation of new types of solitons [23]. It should be noted that the model describing light propagation in a finite nematic liquid crystal without a static biasing electric field is mathematically equivalent to the model describing light propagation in a thermal medium; thus, boundary effects in such crystals may also considerably affect the propagation dynamics of light beams [24]. Soliton formation in different nonlocal materials, including thermal media, with a periodic refractive index modulations have been investigated extensively during the last years [25-28]. However, the effect of localized defects - for example regions with locally increased refractive index - on propagation of laser radiation in strongly nonlocal thermal medium has not been addressed yet.

In this paper we show that boundary-mediated soliton deflection in thermal nonlinear media can be used for controllable soliton ejection from localized trapping potentials created by regions with locally increased refractive index displaced from the center of sample. We show that while in uniform thermal media light beams launched off-center always experience deflection toward the center of the sample due to specific boundary-dependent nonlinear refractive index profile, in the presence of trapping defects one can achieve stationary soliton propagation provided that the soliton power is below a critical value, which depends on the displacement of trapping defect from the center of the sample and on the defect strength. When the soliton power exceeds the critical value, solitons are ejected away from the trapping potential. Note that soliton ejection from trapping potentials studied recently in liquid crystals [27,28] is mediated by input phase tilts, in sharp contrast to the scheme that we consider here.

We consider the propagation of a laser beam along the $\xi$ axis of a thermal medium occupying the region $-L/2 \leq \eta \leq +L/2$ described by a system of coupled equations for the dimensionless field amplitude $q$ and the nonlinear contribution to the refractive index $n$ that is proportional to the local temperature variation (for full details of the derivation, see [7,8]):

$$i\frac{\partial q}{\partial \xi} = -\frac{1}{2}\frac{\partial^2 q}{\partial \eta^2} - qn - pV(\eta - d)q,$$
$$\frac{\partial^2 n}{\partial \eta^2} = -|q|^2,$$
(1)

Here $q = (k_0^2 x_0^4 \alpha |\beta/\kappa n_0|)^{1/2} A$ is the dimensionless light field amplitude; $n = k_0^2 x_0^2 \delta n / n_0$ is proportional to the nonlinear change $\delta n$ of the refractive index $n_0$; $\alpha, \beta, \kappa$ are the optical absorption coefficient, the thermo-optic coefficient, and the thermal conductivity coefficient, respectively. In a typical thermal sample made of lead glass one has $n_0 = 1.8$, $\beta \sim 10^{-5}$ K$^{-1}$, $\alpha \sim 0.01$ cm$^{-1}$ and $\kappa \sim 1$ W m$^{-1}$ K$^{-1}$ [7,8]. The transverse and longitudinal coordinate $\eta, \xi$ are scaled to the beam width $x_0$ and to the diffraction length $k_0 x_0^2$, respectively; the function $V(\eta)$ describes the profile of refractive index defect centered around $\eta = d$, and the parameter $p$ describes the strength of the defect. Further we consider defects described by $V(\eta) = \exp(-\eta^2 / w_\eta^2)$ with $w_\eta = 1$, but the results do not change qualitatively for other defect shapes. Such localized refractive defects in thermal media may be imprinted technologically or induced via the electro-optic effect upon illumination of a sample with a suitable high-power narrow beam at the wavelength at which absorption is sufficiently low. We set $L = 40$ in accordance with typical experimental conditions. It should be noted that the system (1) can be recast into a single integro-differential equation:

$$i\frac{\partial q}{\partial \xi} = -\frac{1}{2}\frac{\partial^2 q}{\partial \eta^2} - pV(\eta - d)q - q\int_{-L/2}^{L/2} G(\eta, \lambda) |q(\lambda)|^2 \, d\lambda, \qquad (2)$$

where $G(\eta, \lambda) = [L/2 + \lambda \mathrm{sgn}(\eta - \lambda)][L/2 - \eta \mathrm{sgn}(\eta - \lambda)] / L$ is the response function of the thermal medium.

In thermal media the conditions imposed at the sample boundaries affect the entire refractive index distribution. Here we assume that both boundaries are thermally stabilized, i.e., $q, n|_{\eta = \pm L/2} = 0$. When the laser beam enters such a medium, it experiences slight absorption resulting in increase and redistribution of temperature in the entire sample. The changes in the temperature are mapped into the nonlinear refractive index profile. Thus, when the beam is launched off-center the temperature redistribution results in the formation of a broad asymmetric refractive index profile with a maximum around the intensity peak. Due to the asymmetry of the induced refractive index profile, a force appears that is pointed toward the sample center. In uniform thermal media such a force may cause periodic soliton center oscillations [16]. However, sufficiently strong external trapping potentials $V(\eta)$ displaced from the center of the sample may arrest such nonlinearity-mediated deflection even for solitons launched close to the sample edge where the deflecting force is most pronounced. The dynamics of the laser beam entering such a medium is thus determined by the competition of nonlinearity-mediated deflection, which becomes more and more pronounced with increase of peak power, and trapping by the potential. One may thus imagine that trapping will occur for beams whose power is below certain critical value, while high-power beams would be ejected from trapping potential.

To show that this is the case we first obtained exact soliton solutions of Eq. (1) in the form $q = w(\eta)\exp(ib\xi)$, where $b$ is the propagation constant. Such solitons can be characterized by their energy flow $U = \int_{-\infty}^{\infty} |q|^2 \, d\eta$. Figure 1(a) shows representative soliton profiles for different propagation constants $b$ (hence, different energy flows $U$). Since nonlinearity-mediated deflection becomes stronger with increase of peak amplitude the position of center of stationary state shifts continuously toward the center of the sample, i.e. to the left wing of trapping potential. The soliton profile that is almost symmetric at $U \to 0$ (in this limit the shape transforms into linear guided mode of potential $V$) becomes notably asymmetric at higher energy flows. In all cases the energy flow was monotonically increasing function of propagation constant. Figure 1(b) shows position of soliton center versus $b$. As one can see the different scenarios take place for $d \geq 3$ and $d < 3$. For large displacements of the trapping potential from the center of the sample there exists a cutoff value $b_{co}$ of propagation constant above which no solitons can be found. The soliton position gradually shifts toward the center of the sample, but the corresponding dependence $\eta_c(b)$ stops at $b = b_{co}$ [see the curve corresponding to $d = 3$ in Fig. 1(b)] that indicates that certain separation between soliton and sample centers always exists. In contrast, when $d < 3$ no cutoff in $b$ was found and with increase of $b$ soliton center shifts smoothly from the center of trap-ping potential into the center of the sample (i.e. $\eta_c \to 0$ as $b \to \infty$). The energy flow of soliton in this case increases indefinitely with $b$.

This difference in soliton behavior at large and small $d$ values can be explained taking into account the fact that for small separations $d$ the soliton, even being located close to the center of the sample, still overlaps effectively with trapping potential whose width $w_\eta \sim d$ in this case. This overlap is sufficient to overcome deflection force that is rather small when soliton is located close to the sample center. Apparently, for large shifts $d \gg w_\eta$ such a balance would not be possible that results in qualitatively different behavior. Remarkably, linear stability analysis indicates that solitons obtained here are stable in the entire existence domain, for all values of the shift $d$ of the external trapping potential.

Next we consider only large displacements $d$ when solitons exist only for propagation constant values $b \leq b_{co}$ and for energy flows below certain maximal (or critical) value $U \leq U_{max}$. At fixed defect strength $p$ both $b_{co}$ and $U_{max}$ are monotonically decreasing functions of displacement $d$ [Figs. 2(a) and 2(c)], which is natural since deflection force also diminishes with decrease of $d$. The dependencies $b_{co}(d)$ and $U_{co}(d)$ in Fig. 2 can be well approximated by the curves $\sim a/d^{1.75}$ with a properly selected interpolation parameter $a$. Notice that both $b_{co}$ and $U_{max}$ diverge around $d = 3$. Stronger trapping potentials can compensate deflection of more powerful beams, which is manifested in monotonic increase of $b_{co}$ and $U_{max}$ with growth of defect strength $p$ [Figs. 2(b) and 2(d)]. The dependencies $b_{co}(p)$ and $U_{co}(p)$ are well approximated by the cubic polynomials $\sim ap + bp^2 + cp^3$.

Next we proceed with the study of the dynamics of soliton ejection. We performed direct numerical integration of Eq. (1) with inputs that do not exactly correspond to solitons, for exam-ple, when $q|_{\xi=0} = A\exp[-(\eta-d)^2]$. Figure 3(a) shows different propagation scenarios of Gaussian beam for different input amplitudes. Consistent with the existence of a maximum energy flow for stationary solitons that can be trapped by the localized potential, we found that upon propagation of input Gaussian beams two distinct propagation scenarios (ejection or trapping) may occur, depending on whether or not the input energy flow exceeds a critical value $U'_{max}$. The critical energy $U'_{max}$ grows with increasing $p$ and diminishes with increasing $d$, similar to the behavior encountered with stationary solutions. For example, for $d=10$ and $p=1$ the critical energy flow amounts to $U'_{max} \approx 1.1$, a value that is a bit smaller than the maximum energy flow of the stationary soliton solution, i.e., $U_{max} \approx 1.7$ for the same set of parameters. This is consistent with the fact that the shape and the input position of the Gaussian beam $q|_{\xi=0} = A\exp[-(\eta-d)^2]$ that we used as input condition differs substantially from the actual asymmetric shape and position of the stationary soliton, whose center is always shifted toward the center of the sample [see Fig. 1(a)]. Hence, in all cases excitation with Gaussian beams yields a critical energy flow that is a bit smaller than the critical energy flow for actual stationary solitons.

In Figure 3(a), when $A=0.9$, the energy flow of the input beam is below the maximal energy flow value $U'_{max}$ corresponding to the selected displacement $d$ and the beam propagates along the defect experiencing only slight oscillations. In contrast, for $A=1$ when $U > U'_{max}$, the nonlinearity-mediated deflection dominates and the beam experiences ejection from the trapping potential. In the course of subsequent propagation such beams oscillate inside the sample periodically returning to the vicinity of launching point.

If a phase tilt $\exp(-i\alpha\eta)$ is imposed on the input conditions, the beams can be ejected from the trapping potential even for $U < U'_{max}$, when nonlinearity-mediated deflection is too weak.

Under such conditions, ejection occurs when the input tilt exceeds certain critical value $\alpha_{cr}$ that monotonically decreases with increase of input amplitude $A$ and that goes to zero for amplitudes where $U \approx U'_{max}$ [Fig. 3(c)]. By changing the input tilt, or by changing the input amplitude, one can control the propagation trajectory and output position of the beam [Fig. 3(b)].

In conclusion, we exposed that in the presence of local refractive index defects embedded in thermal nonlinear media one may induce either trapping of light beams launched close to the sample boundary or rather their transverse ejection. The process takes place in a controllable manner. Importantly, for a given material and geometrical setting, the control parameter is the input light power, opening up new possibilities for self-routing and self-steering of light beams.

# References

1. A. W. Snyder and D. J. Mitchell, "Accessible solitons," Science **276**, 1538 (1997).

2. W. Krolikowski, O. Bang, N. I. Nikolov, D. Neshev, J. Wyller, J. J. Rasmussen, and D. Edmundson, "Modulational instability, solitons and beam propagation in spatially nonlocal nonlinear media," J. Opt. B **6**, S288 (2004).

3. C. Conti, M. C. Conti, M. Peccianti, and G. Assanto, "Route to nonlocality and observation of accessible solitons," Phys. Rev. Lett. **91**, 073901 (2003).

4. C. Conti, M. Peccianti, and G. Assanto, "Observation of optical spatial solitons in a highly nonlocal medium," Phys. Rev. Lett. **92**, 113902 (2004).

5. D. Suter and T. Blasberg, "Stabilization of transverse solitary waves by a nonlocal response of the nonlinear medium," Phys. Rev. A **48**, 4583 (1993).

6. S. Skupin, M. Saffman, and W. Krolikowski, "Nonlocal stabilization of nonlinear beams in a self-focusing atomic vapor," Phys. Rev. Lett. **98**, 263902 (2007).

7. C. Rotschild, O. Cohen, O. Manela, M. Segev, and T. Carmon, "Solitons in nonlinear media with an infinite range of nonlocality: first observation of coherent elliptic solitons and of vortex-ring solitons," Phys. Rev. Lett. **95**, 213904 (2005).

8. C. Rotschild, B. Alfassi, O. Cohen and M. Segev, "Long-range interactions between optical solitons," Nature Physics **2**, 769 (2006).

9. E. Gilad, J. von Hardenberg, A. Provenzale, M. Shachak, and E. Meron, "Ecosystem engineers: From pattern formation to habitat creation," Phys. Rev. Lett. **93**, 098105 (2004).

10. M. G. Clerc, D. Escaff, and V. M. Kenkre, "Patterns and localized structures in population dynamics," Phys. Rev. E **72**, 056217 (2005).

11. V. Volpert and S. Petrovskii, "Reaction-diffusion waves in biology," Phys. Life. Rev. **6**, 267 (2009).

12. S. K. Turitsyn, "Spatial dispersion of nonlinearity and stability of multidimensional solitons," Theor. Math. Phys. **64**, 226 (1985).

13. V. M. Perez-Garcia, V. V. Konotop, and J. J. Garcia-Ripoll, "Dynamics of quasicollapse in nonlinear Schrödinger systems with nonlocal interactions," Phys. Rev. E **62**, 4300 (2000).

14. O. Bang, W. Krolikowski, J. Wyller, and J. J. Rasmussen, "Collapse arrest and soliton stabilization in nonlocal nonlinear media," Phys. Rev. E **66**, 046619 (2002).

15. M. Peccianti, K. A. Brzdakiewicz, and G. Assanto, "Nonlocal spatial soliton interactions in nematic liquid crystals," Opt. Lett. **27**, 1460 (2002).

16. B. Alfassi, C. Rotschild, O. Manela, M. Segev, and D. N. Christodoulides, "Boundary force effects exerted on solitons in highly nonlocal nonlinear media," Opt. Lett. **32**, 154 (2007).

17. A. Alberucci, M. Peccianti, and G. Assanto, "Nonlinear bouncing of nonlocal spatial solitons at the boundaries," Opt. Lett. **32**, 2795 (2007).

18. D. Buccoliero, A. S. Desyatnikov, W. Krolikowski, and Y. Kivshar, "Boundary effects on the dynamics of higher-order optical spatial solitons in nonlocal thermal media," J. Opt. A: Pure Appl. Opt. **11**, 094014 (2009).

19. F. Ye, Y. V. Kartashov, B. Hu, and L. Torner, "Power-dependent soliton steering in thermal nonlinear media," Opt. Lett. **34**, 2658 (2009).

20. B. Alfassi, C. Rotschild, O. Manela, M. Segev, and D. N. Christodoulides, "Nonlocal surface-wave solitons," Phys. Rev. Lett. **98**, 213901 (2007).

21. Y. V. Kartashov, F. Ye, V. A. Vysloukh, and L. Torner, "Surface waves in defocusing thermal media," Opt. Lett. **32**, 2260 (2007).

22. F. Ye, Y. V. Kartashov, and L. Torner, "Nonlocal surface dipoles and vortices," Phys. Rev. A **77**, 033829 (2008).

23. F. Ye, Y. V. Kartashov, B. Hu, and L. Torner, "Twin-vortex solitons in nonlocal nonlinear media," Opt. Lett. **35**, 628 (2010).

24. A. Alberucci, G. Assanto, D. Buccoliero, A. S. Desyatnikov, T. R. Marchant, and N. F. Smyth, "Modulation analysis of boundary-induced motion of optical solitary waves in nematic liquid crystal," Phys. Rev. A **79**, 043816 (2009).

25. Z. Xu, Y. V. Kartashov, and L. Torner, "Soliton mobility in nonlocal optical lattices," Phys. Rev. Lett. **95**, 113901(2005).

26. Y. V. Kartashov, L. Torner, and V. A. Vysloukh, "Lattice-supported surface solitons in nonlocal nonlinear media," Opt. Lett. **31**, 2595(2006).

27. N. K. Efremidis, "Nonlocal lattice solitons in thermal media," Phys. Rev. A **77**, 063824 (2008).

28. P. D. Rasmussen, F. H. Bennet, D. N. Neshev, A. A. Sukhorukov, C. R. Rosberg, W. Krolikowski, O. Bang, and Y. S. Kivshar, "Observation of two-dimensional nonlocal gap solitons," Opt. Lett. **34**, 295 (2009).

29. G. Assanto, A. A. Minzoni, M. Peccianti, and N. F. Smyth, "Optical solitary waves escaping a wide trapping potential in nematic liquid crystals: Modulation theory," Phys. Rev. A **79**, 033837 (2009).

30. M. Peccianti, A. Dyadyusha, M. Kaczmarek, and G. Assanto, "Escaping solitons from a trapping potential," Phys. Rev. Lett. **101**, 153902 (2008).

# Figure captions

Figure 1.  (a) Profiles of solitons with $b=3$ and $b=13$ at $p=1$, $d=10$. Shaded region indicates trapping potential. (b) Integral soliton center versus $b$ at $p=1$. The cutoff value of propagation constant at $d=3$ is given by $b=116.9$. All quantities are plotted in arbitrary dimensionless units.

Figure 2.  Propagation constant cutoff versus $d$ at $p=1$ (a) and versus $p$ at $d=10$ (b). Maximal energy flow of trapped soliton versus $d$ at $p=1$ (c) and versus $p$ at $d=10$ (d). All quantities are plotted in arbitrary dimensionless units.

Figure 3.  (color online) Dynamics of propagation of Gaussian beams with (a) different amplitudes at $\alpha=0$ and (b) with different input tilts at $A=0.93$ for $d=10$, $p=1$. The intensity distributions corresponding to different amplitudes or tilts are superimposed. (c) Critical tilt versus input amplitude at $d=10$, $p=1$. All quantities are plotted in arbitrary dimensionless units.

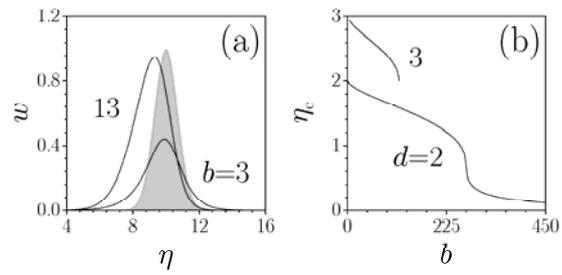

Figure 1.    (a) Profiles of solitons with $b=3$ and $b=13$ at $p=1$, $d=10$. Shaded region indicates trapping potential. (b) Integral soliton center versus $b$ at $p=1$. The cutoff value of propagation constant at $d=3$ is given by $b=116.9$. All quantities are plotted in arbitrary dimensionless units.

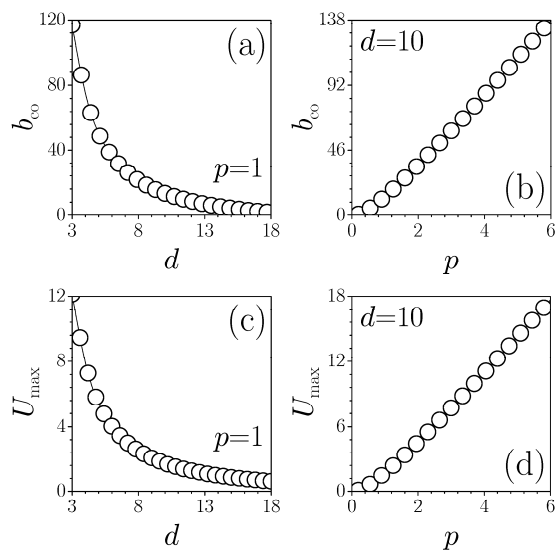

Figure 2. Propagation constant cutoff versus $d$ at $p=1$ (a) and versus $p$ at $d=10$ (b). Maximal energy flow of trapped soliton versus $d$ at $p=1$ (c) and versus $p$ at $d=10$ (d). All quantities are plotted in arbitrary dimensionless units.

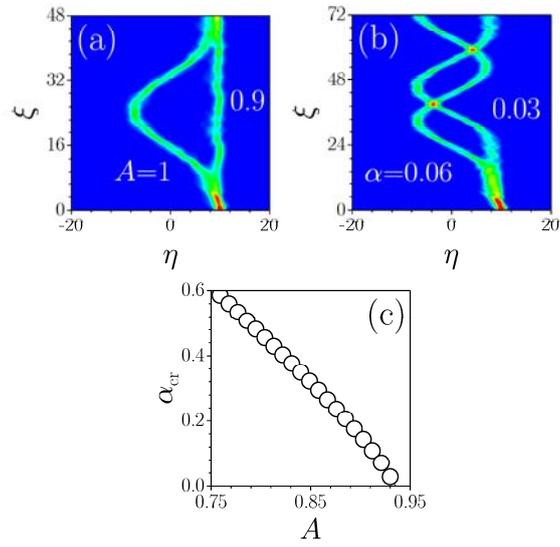

Figure 3. (color online) Dynamics of propagation of Gaussian beams with (a) different amplitudes at $\alpha = 0$ and (b) with different input tilts at $A = 0.93$ for $d = 10$, $p = 1$. The intensity distributions corresponding to different amplitudes or tilts are superimposed. (c) Critical tilt versus input amplitude at $d = 10$, $p = 1$. All quantities are plotted in arbitrary dimensionless units.